\documentclass[
  aps,
  prl,
  reprint,
  superscriptaddress,
  nofootinbib,
  longbibliography
]{revtex4-2}

\usepackage{listings}

\lstdefinestyle{matlabcompact}{
  language=Matlab,
  basicstyle=\ttfamily\fontsize{7.5}{7.2}\selectfont,
  numbers=none,
  frame=none,
  breaklines=true,
  breakatwhitespace=false,
  breakindent=0pt,
  columns=fullflexible,
  keepspaces=true,
  showstringspaces=false,
  tabsize=2,
  linewidth=\columnwidth,
  xleftmargin=0pt,
  xrightmargin=0pt,
  resetmargins=true,
  aboveskip=2pt,
  belowskip=2pt
}
\usepackage{amsmath,amssymb}
\usepackage{array}[=2016-10-06]
\usepackage{booktabs,longtable}
\usepackage{graphicx}
\usepackage{microtype}
\usepackage{hyperref}
 \hypersetup{colorlinks,linkcolor={blue},citecolor={red},urlcolor={blue}}
\newtheorem{lemma}{Lemma}

\newcommand{\bbC}{\mathbb{C}}
\newtheorem{theorem}{Theorem}
\newcommand{\bra}[1]{\langle#1|}
	\newcommand{\ket}[1]{|#1\rangle}

	\newcommand{\braket}[2]{\langle#1|#2\rangle}

\newcommand{\rank}{\operatorname{rank}}

\begin{document}

\title{A Minimum-Cardinality Genuinely Unextendible Product Basis in Three Qutrits}

\author{Fei Shi}
   \email[]{shif26@mail.sysu.edu.cn}
	\affiliation{Institute of Quantum Computing and Software, School of Computer Science and Engineering, Sun Yat-sen University, Guangzhou 510006, China}	

\author{Ge Bai}
   \email[]{gebai@hkust-gz.edu.cn}
\affiliation{Thrust of Artificial Intelligence, Information Hub, The Hong Kong University of Science and Technology (Guangzhou), Guangzhou 511453, China}

\author{Xiande Zhang}
   \email[]{drzhangx@ustc.edu.cn}
\affiliation{School of Mathematical Sciences,
		University of Science and Technology of China, Hefei, 230026, China; and with Hefei National Laboratory, University of Science and Technology of China, Hefei 230088, China}
        
\author{Qi Zhao}
\email[]{zhaoqi@cs.hku.hk}
 \affiliation{QICI Quantum Information and Computation Initiative, School of Computing and Data Science,
The University of Hong Kong, Pokfulam Road, Hong Kong SAR, China}

% \author{Giulio Chiribella}
% \email[]{giulio@cs.hku.hk}
% \affiliation{QICI Quantum Information and Computation Initiative, School of Computing and Data Science,
% The University of Hong Kong, Pokfulam Road, Hong Kong SAR, China}
% \affiliation{Department of Computer Science, Parks Road, Oxford, OX1 3QD, United Kingdom}
%  \affiliation{Perimeter Institute for Theoretical Physics, Waterloo, Ontario N2L 2Y5, Canada}

\author{Lvzhou Li}
\email[]{lilvzh@mail.sysu.edu.cn} 
	\affiliation{Institute of Quantum Computing and Software, School of Computer Science and Engineering, Sun Yat-sen University, Guangzhou 510006, China}

\begin{abstract}
It has remained an open question whether a genuinely unextendible
product basis (GUPB) exists. We resolve this problem by constructing an
explicit three-qutrit GUPB of cardinality fourteen in the smallest
tripartite Hilbert space in which a GUPB can exist. Together with the
nonexistence of three-qutrit GUPBs of cardinality less than fourteen,
our construction proves that fourteen is the minimum cardinality. A
padding procedure further extends the construction to all tripartite
systems whose local dimensions are at least three. As applications,
the normalized projector onto the thirteen-dimensional orthogonal
complement of the three-qutrit GUPB is positive under partial
transposition and bound entangled across every bipartition, while the
GUPB exhibits strong quantum nonlocality without entanglement.
\end{abstract}

\maketitle

\textit{\textbf{Introduction.}}\textbf{---}
An unextendible product basis (UPB) is a set of
orthogonal product states whose orthogonal
complement contains no product state \cite{Bennett1999}. The normalized projector onto the
orthogonal complement of a UPB is a positive-partial-transpose (PPT)
entangled state and is therefore bound entangled: no pure entanglement
can be distilled from it by local operations and classical
communication \cite{Bennett1999,Horodecki1998,DiVincenzo2003}. UPBs have also been
used to construct Bell inequalities that admit no quantum violation
\cite{Augusiak2011,Augusiak2012,fritz2013local}, providing insight into
the foundations of quantum theory. The construction of UPBs has also attracted considerable attention
\cite{AL01,Fen06,Chen2013The,Joh13,Johnston2014The,
shi2020unextendible}.

A genuinely unextendible product basis (GUPB) is a set of orthogonal product states whose orthogonal complement contains no
biproduct state
\cite{DemianowiczAugusiak2018}. Equivalently, a GUPB is a UPB across
every bipartition. General lower bounds on the cardinality of GUPBs
have been established \cite{Demianowicz2022,Shi2023}. Orthogonal
product sets that cannot be completed to full product bases across
every bipartition are also known \cite{Shi2022}; however,
uncompletability is weaker than genuine unextendibility and does not
exclude all biproduct vectors from the orthogonal complement.
Consequently, the existence of orthogonal GUPBs remained open.

A bipartite UPB cannot exist in $\bbC^2\otimes\bbC^n$ \cite{Bennett1999,DiVincenzo2003}. Hence, if one local subsystem has dimension $2$, a GUPB can not exist. The
three-qutrit space $(\bbC^3)^{\otimes3}$ is therefore the smallest
tripartite setting in which a GUPB can exist. Previous lower bounds
imply that a three-qutrit GUPB must contain at least $13$ states
\cite{Demianowicz2022,Shi2023}, while cardinality $13$ has recently
been excluded \cite{Demianowicz2026}. Thus, the minimum possible
cardinality is at least $14$.

In this Letter, we resolve the existence problem by constructing a
three-qutrit GUPB of cardinality $14$, thereby attaining the lower
bound and determining the minimum cardinality exactly. We further
extend the construction to every tripartite Hilbert space
$\bbC^{d_A}\otimes\bbC^{d_B}\otimes\bbC^{d_C}$ with
$3\leq d_A\leq d_B\leq d_C$. Finally, we show that the normalized
projector onto the orthogonal complement of the constructed GUPB is
PPT and bound entangled across every bipartition, and that the GUPB
exhibits strong quantum nonlocality without entanglement.

\textit{\textbf{A three-qutrit GUPB of cardinality 14.}}\textbf{---} Throughout this Letter, we work with unnormalized vectors, since
normalization affects neither orthogonality nor rank. A nonzero vector
$z=(z_1,z_2,z_3)\in\bbC^3$ represents, up to normalization, the pure state
\[
\ket{z}=z_1\ket{0}+z_2\ket{1}+z_3\ket{2},
\]
where $\{\ket{0},\ket{1},\ket{2}\}$ is the computational basis of
$\bbC^3$. We also use the notation $[n]=\{1,2,\ldots,n\}$. 

  \begin{figure*}[t]
\centering
\includegraphics[width=0.94\textwidth]{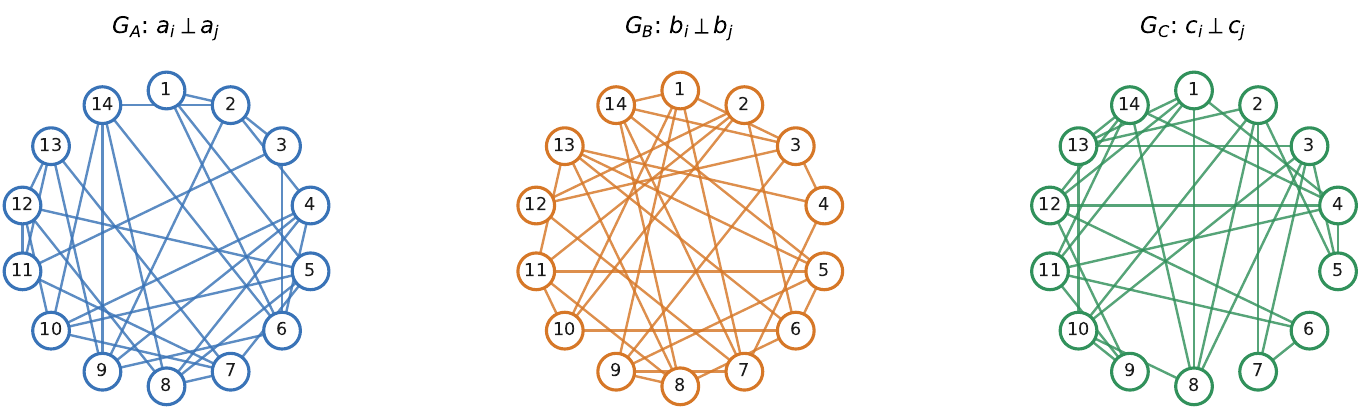}
\caption{\label{fig:local-graphs}Local orthogonality graphs of the construction. Vertex $i$ labels the
$i$-th product state in Table~\ref{tab:construction}, and an edge
$(i,j)$ in $G_X$ indicates that the corresponding local vectors at
party $X$ are orthogonal. All three panels use the same circular
ordering of the vertices, and their edge sets together cover
$K_{14}$. The graphs $G_A$ and $G_B$ are isomorphic under the
permutation $\pi$ defined in Eq.~\eqref{eq:ab-permutation}.  }
\end{figure*}

For a family of nonzero vectors
$\{\ket{\psi_i}\}_{i\in[n]}$, its \emph{orthogonality graph} is the
graph $G=(V,E)$ with vertex set $V=[n]$ and edge set
\begin{equation}
E=\bigl\{(i,j):1\leq i<j\leq n,\ 
\braket{\psi_i}{\psi_j}=0\bigr\}.
\label{eq:orthogonality-graph}
\end{equation}
For two graphs $G_1=(V,E_1)$ and $G_2=(V,E_2)$ on the same vertex
set, their union is defined as
$G_1\cup G_2=(V,E_1\cup E_2)$. We denote by $K_n$ the complete graph on $n$ vertices, in which every
pair of distinct vertices is adjacent. Conversely, an \emph{orthogonal representation} of a graph
$G=(V,E)$ in $\bbC^d$ assigns a nonzero vector
$\ket{\psi_i}\in\bbC^d$ to each vertex $i\in V$ such that
$\braket{\psi_i}{\psi_j}=0$ whenever $(i,j)\in E$. The representation
is called \emph{faithful} if
$\braket{\psi_i}{\psi_j}=0$ holds if and only if
$(i,j)\in E$.

Consider a family of tripartite product states
$\{\ket{x_i}\otimes\ket{y_i}\otimes\ket{z_i}\}_{i\in[n]}$, and let
$G_A$, $G_B$, and $G_C$ be the orthogonality graphs of
$\{\ket{x_i}\}_{i\in[n]}$, $\{\ket{y_i}\}_{i\in[n]}$, and
$\{\ket{z_i}\}_{i\in[n]}$, respectively. Since
$\braket{x_i}{x_j}\braket{y_i}{y_j}\braket{z_i}{z_j}=0$ if and only
if at least one of the three local inner products vanishes, the product
states are mutually orthogonal if and only if
$G_A\cup G_B\cup G_C=K_n$.

\begin{table}[t]
\caption{Unnormalized real local vectors for
the three-qutrit GUPB of
cardinality $14$. 
}\label{tab:construction}
\centering
\setlength{\tabcolsep}{13pt}
\renewcommand{\arraystretch}{1.04}
\begin{tabular}{c c c c}
\toprule
$i$ & $a_i$ & $b_i$ & $c_i$\\
\midrule
1 & $(0,-1,2)$ & $(0,0,1)$ & $(1,0,0)$ \\
2 & $(1,0,0)$ & $(1,-1,-2)$ & $(2,1,-1)$ \\
3 & $(0,2,-1)$ & $(1,0,0)$ & $(2,1,-1)$ \\
4 & $(0,1,0)$ & $(0,-1,2)$ & $(0,1,0)$ \\
5 & $(1,2,1)$ & $(1,0,-1)$ & $(1,0,2)$ \\
6 & $(1,0,0)$ & $(1,-1,1)$ & $(1,3,0)$ \\
7 & $(1,-1,1)$ & $(1,0,0)$ & $(3,-1,5)$ \\
8 & $(1,0,-1)$ & $(1,0,-1)$ & $(0,1,1)$ \\
9 & $(0,0,1)$ & $(0,1,0)$ & $(1,1,0)$ \\
10 & $(1,0,-1)$ & $(1,1,0)$ & $(-1,1,-1)$ \\
11 & $(1,-1,-2)$ & $(1,-1,1)$ & $(0,0,1)$ \\
12 & $(1,-1,1)$ & $(0,2,-1)$ & $(0,0,1)$ \\
13 & $(1,1,0)$ & $(1,2,1)$ & $(0,1,1)$ \\
14 & $(0,1,0)$ & $(0,1,0)$ & $(1,0,0)$ \\
\bottomrule
\end{tabular}
\end{table}

Guided by the graph-theoretic approach developed in our earlier work
\cite{Shi2023}, we performed a computer-assisted search for triples of
graphs $G_A$, $G_B$, and $G_C$ on fourteen vertices whose union is
$K_{14}$ and that admit compatible faithful orthogonal representations
in $\bbC^3$. To
reduce the search space, we imposed the symmetry $G_A\cong G_B$ and
tested each candidate against the unextendibility criterion across all
three bipartitions. A successful representation was subsequently
converted into the integer vectors listed in
Table~\ref{tab:construction}. The computer-assisted search was used only to discover the
construction; its mutual orthogonality and genuine unextendibility are
established below by exact calculations. This procedure leads to our
main result.

\begin{theorem}
The product states
$\{\ket{a_i}\otimes\ket{b_i}\otimes\ket{c_i}\}_{i\in[14]}$
listed in Table~\ref{tab:construction} form a three-qutrit GUPB of
cardinality $14$, which is the minimum possible cardinality of any
three-qutrit GUPB.
\end{theorem}

We first establish the mutual orthogonality of the product states
$\{\ket{a_i}\otimes\ket{b_i}\otimes\ket{c_i}\}_{i\in[14]}$.
Let $G_A=([14],E_A)$, $G_B=([14],E_B)$, and
$G_C=([14],E_C)$ denote the orthogonality graphs of
$\{\ket{a_i}\}_{i\in[14]}$, $\{\ket{b_i}\}_{i\in[14]}$, and
$\{\ket{c_i}\}_{i\in[14]}$, respectively. These graphs are shown in
Fig.~\ref{fig:local-graphs}.
The graphs $G_A$ and $G_B$ are isomorphic. Indeed, define the
permutation $\pi$ of the vertex set $[14]$ by
\begin{equation}
\pi=(1\,4\,9)(2\,3\,12\,11)(5\,13\,10\,8)(6\,7),
\qquad \pi(14)=14.
\label{eq:ab-permutation}
\end{equation}
By construction, $\ket{b_{\pi(i)}}=\ket{a_i}$ for every  $i\in[14]$.  The edge sets of $G_A$, $G_B$, and $G_C$ satisfy
$|E_A|=|E_B|=32$, $|E_C|=31$.
Among the edges of $K_{14}$, the edge $(11,13)$ belongs to $E_A$ and
$E_B$ but not to $E_C$, the edge $(2,10)$ belongs to $E_B$ and $E_C$
but not to $E_A$, and the edge $(8,14)$ belongs to all three edge
sets. Every other edge $(i,j)$ of $K_{14}$ belongs to exactly one of
$E_A$, $E_B$, and $E_C$. Consequently,
$G_A\cup G_B\cup G_C=K_{14}$.
Therefore, the product states $\left\{
\ket{a_i}\otimes\ket{b_i}\otimes\ket{c_i}
\right\}_{i\in[14]}$
are mutually orthogonal.

To establish the GUPB property, it suffices to show that the three
families
\[
\begin{aligned}
&\left\{
\ket{a_i}\otimes
\bigl(\ket{b_i}\otimes\ket{c_i}\bigr)
\right\}_{i\in[14]},\\
&\left\{
\ket{b_i}\otimes
\bigl(\ket{a_i}\otimes\ket{c_i}\bigr)
\right\}_{i\in[14]},\\
&\left\{
\ket{c_i}\otimes
\bigl(\ket{a_i}\otimes\ket{b_i}\bigr)
\right\}_{i\in[14]}
\end{aligned}
\]
are bipartite UPBs in $\bbC^3\otimes\bbC^9$ across the bipartitions
$A|BC$, $B|AC$, and $C|AB$, respectively. Here
$\ket{a_i},\ket{b_i},\ket{c_i}\in\bbC^3$, and 
$\ket{b_i}\otimes\ket{c_i}$,
$\ket{a_i}\otimes\ket{c_i}$,
$\ket{a_i}\otimes\ket{b_i}\in \bbC^9$. According to Ref.~\cite{Bennett1999}, we have the following lemma.

\begin{lemma}\label{lemma:criterion}
Let $
\{\ket{x_i}\otimes\ket{y_i}\}_{i\in[14]}
\subset \bbC^3\otimes\bbC^9$
be a set of orthogonal product states, where
$\ket{x_i}\in\bbC^3$ and $\ket{y_i}\in\bbC^9$. Then $\{\ket{x_i}\otimes\ket{y_i}\}_{i\in[14]}$ is a
UPB if and only if, for every subset $J\subseteq[14]$,
\begin{equation}
\operatorname{rank}\{\ket{x_j}:j\in J\}=3
\ \text{or} \
\operatorname{rank}\{\ket{y_j}:j\in J^c\}=9.
\label{eq:criterion}
\end{equation}
\end{lemma}

Indeed, suppose that there exists a subset $J\subseteq[14]$ such that $\rank\{\ket{x_j}:j\in J\}<3
$ and $
\rank\{\ket{y_j}:j\in J^c\}<9.
$
Then there exist nonzero vectors $\ket{u}\in\bbC^3$ and
$\ket{v}\in\bbC^9$ satisfying $\braket{u}{x_j}=0$ for all $j\in J$, $\braket{v}{y_j}=0$ for all $j\in J^c$.
Consequently, for every $i\in[14]$,
$\braket{u\otimes v}{x_i\otimes y_i}=0$. Conversely, suppose that there exists a nonzero product state
$\ket{u}\otimes\ket{v}$ orthogonal to every
$\ket{x_i}\otimes\ket{y_i}$. Define $J=\{i\in[14]:\braket{u}{x_i}=0\}$.
Then $\operatorname{span}\{\ket{x_j}:j\in J\}\subseteq u^\perp$,
and hence 
$\rank\{\ket{x_j}:j\in J\}<3$. For every $j\in J^c$, we have $\braket{u}{x_j}\neq 0$. Since $
\braket{u}{x_j}\braket{v}{y_j}=0$,
it follows that $\braket{v}{y_j}=0$. Therefore,
$\operatorname{span}\{\ket{y_j}:j\in J^c\}\subseteq v^\perp$, which implies $\rank\{\ket{y_j}: j\in J^c\}<9$.

% For $A|BC$, Eq.~\eqref{eq:criterion} uses
% $x_i=a_i\in\CC^3$ and $y_i=b_i\otimes c_i\in\CC^9$; the other cuts are
% cyclic.  Notice that the vector in the two-party block is arbitrary in
% $\CC^9$ and need not factor further, so this criterion tests genuine
% rather than merely fully product unextendibility.  Equivalently, on
% each cut one must establish the finite family of alternatives
% \begin{equation}
%  r_X(J)=3\quad\text{or}\quad r_{\bar X}(J^c)=9
%  \qquad\text{for every }J\subseteq[14],
%  \label{eq:rank-alternative}
% \end{equation}
% where $r_X(J)$ and $r_{\bar X}(J^c)$ are the two indicated span
% dimensions.  This form makes the certificate independently
% checkable: no numerical optimization or genericity assumption enters
% the proof.

\begin{table}[t]
\caption{\label{tab:app-deficient}Locally deficient subsets by cardinality.}
\centering
\setlength{\tabcolsep}{6pt}
\renewcommand{\arraystretch}{1.04}
\begin{tabular}{c| c c c c c c |c}
\toprule
bipartition & 0 & 1 & 2 & 3 & 4 & 5 & Total\\
\midrule
$A|BC$ & 1 & 14 & 91 & 82 & 34 & 6 & 228\\
$B|AC$ & 1 & 14 & 91 & 82 & 34 & 6 & 228\\
$C|AB$ & 1 & 14 & 91 & 78 & 32 & 6 & 222\\
\bottomrule
\end{tabular}
\end{table}

\begin{table}[t]
\caption{\label{tab:inclusion-maximal}Inclusion-maximal deficient subsets by cardinality.}
\setlength{\tabcolsep}{6pt}
\renewcommand{\arraystretch}{1.2}
\centering
\begin{tabular}{c|c c c c|c}
\toprule
\text{bipartition} & 2&3&4&5&\text{Total}\\ \midrule
$A|BC$&7&6&4&6&23\\
$B|AC$&7&6&4&6&23\\
$C|AB$&9&10&2&6&27\\
\bottomrule
\end{tabular}
\end{table}

\begin{table*}[t]
\caption{\label{tab:app-determinants}
Complete determinant certificates. For each bipartition, $M$ is an
inclusion-maximal deficient subset, $P\subseteq M^c$ contains nine
selected pair-vector indices, and $\det$ is the corresponding
nonzero determinant.}
\centering
\fontsize{6.1}{6.9}\selectfont
\setlength{\tabcolsep}{1.6pt}
\renewcommand{\arraystretch}{1}

% ==================== A|BC ====================
\begin{minipage}[t]{0.325\textwidth}
\vspace{0pt}
\centering
\begin{tabular}{@{}l l r@{}}
\multicolumn{3}{c}{\textbf{(a) bipartition $A|BC$}}\\
\toprule
$M$ & $P\subseteq M^c$ & $\det$\\
\midrule
$\{1,5\}$ & $\{2,3,4,6,7,8,9,10,11\}$ & $-918$\\
$\{1,11\}$ & $\{2,3,4,5,6,7,8,9,10\}$ & $-198$\\
$\{1,13\}$ & $\{2,3,4,5,6,7,8,9,10\}$ & $-198$\\
$\{3,5\}$ & $\{1,2,4,6,7,8,9,10,11\}$ & $-74$\\
$\{3,11\}$ & $\{1,2,4,5,6,7,8,9,10\}$ & $-90$\\
$\{5,9\}$ & $\{1,2,3,4,6,7,8,10,11\}$ & $112$\\
$\{9,13\}$ & $\{1,2,3,4,5,6,7,8,10\}$ & $36$\\
$\{1,2,6\}$ & $\{3,4,5,7,8,9,10,11,12\}$ & $129$\\
$\{2,3,6\}$ & $\{1,4,5,7,8,9,10,11,12\}$ & $-47$\\
$\{2,5,6\}$ & $\{1,3,4,7,8,9,10,11,12\}$ & $-129$\\
$\{2,6,11\}$ & $\{1,3,4,5,7,8,9,10,12\}$ & $282$\\
$\{3,8,10\}$ & $\{1,2,4,5,6,7,9,11,12\}$ & $-31$\\
$\{4,11,14\}$ & $\{1,2,3,5,6,7,8,9,10\}$ & $432$\\
$\{2,6,7,12\}$ & $\{1,3,4,5,8,9,10,11,13\}$ & $-272$\\
$\{3,7,12,13\}$ & $\{1,2,4,5,6,8,9,10,11\}$ & $54$\\
$\{4,8,10,14\}$ & $\{1,2,3,5,6,7,9,11,12\}$ & $-26$\\
$\{7,9,11,12\}$ & $\{1,2,3,4,5,6,8,10,13\}$ & $108$\\
$\{1,3,4,9,14\}$ & $\{2,5,6,7,8,10,11,12,13\}$ & $2160$\\
$\{1,7,8,10,12\}$ & $\{2,3,4,5,6,9,11,13,14\}$ & $-540$\\
$\{2,4,6,13,14\}$ & $\{1,3,5,7,8,9,10,11,12\}$ & $-52$\\
$\{2,6,8,9,10\}$ & $\{1,3,4,5,7,11,12,13,14\}$ & $-130$\\
$\{4,5,7,12,14\}$ & $\{1,2,3,6,8,9,10,11,13\}$ & $648$\\
$\{5,8,10,11,13\}$ & $\{1,2,3,4,6,7,9,12,14\}$ & $350$\\
\bottomrule
\end{tabular}
\end{minipage}
\hfill
% ==================== B|AC ====================
\begin{minipage}[t]{0.325\textwidth}
\vspace{0pt}
\centering
\begin{tabular}{@{}l l r@{}}
\multicolumn{3}{c}{\textbf{(b) bipartition $B|AC$}}\\
\toprule
$M$ & $P\subseteq M^c$ & $\det$\\
\midrule
$\{1,10\}$ & $\{2,3,4,5,6,7,8,9,11\}$ & $-2522$\\
$\{1,13\}$ & $\{2,3,4,5,6,7,8,9,10\}$ & $6252$\\
$\{2,4\}$ & $\{1,3,5,6,7,8,9,10,11\}$ & $-594$\\
$\{2,12\}$ & $\{1,3,4,5,6,7,8,9,10\}$ & $2430$\\
$\{4,10\}$ & $\{1,2,3,5,6,7,8,9,11\}$ & $-256$\\
$\{4,13\}$ & $\{1,2,3,5,6,7,8,9,10\}$ & $2016$\\
$\{12,13\}$ & $\{1,2,3,4,5,6,7,8,9\}$ & $3056$\\
$\{2,3,7\}$ & $\{1,4,5,6,8,9,10,11,12\}$ & $-66$\\
$\{2,9,14\}$ & $\{1,3,4,5,6,7,8,10,11\}$ & $-1656$\\
$\{3,4,7\}$ & $\{1,2,5,6,8,9,10,11,13\}$ & $-22$\\
$\{3,7,12\}$ & $\{1,2,4,5,6,8,9,10,11\}$ & $-22$\\
$\{3,7,13\}$ & $\{1,2,4,5,6,8,9,10,11\}$ & $-22$\\
$\{5,8,12\}$ & $\{1,2,3,4,6,7,9,10,11\}$ & $-685$\\
$\{1,2,6,11\}$ & $\{3,4,5,7,8,9,10,12,13\}$ & $522$\\
$\{3,6,7,11\}$ & $\{1,2,4,5,8,9,10,12,13\}$ & $-170$\\
$\{5,8,9,14\}$ & $\{1,2,3,4,6,7,10,11,12\}$ & $-273$\\
$\{6,10,11,12\}$ & $\{1,2,3,4,5,7,8,9,13\}$ & $-1044$\\
$\{1,3,5,7,8\}$ & $\{2,4,6,9,10,11,12,13,14\}$ & $-54$\\
$\{1,4,9,12,14\}$ & $\{2,3,5,6,7,8,10,11,13\}$ & $-6300$\\
$\{2,5,8,10,13\}$ & $\{1,3,4,6,7,9,11,12,14\}$ & $120$\\
$\{3,7,9,10,14\}$ & $\{1,2,4,5,6,8,11,12,13\}$ & $-135$\\
$\{4,5,6,8,11\}$ & $\{1,2,3,7,9,10,12,13,14\}$ & $-104$\\
$\{6,9,11,13,14\}$ & $\{1,2,3,4,5,7,8,10,12\}$ & $1560$\\
\bottomrule
\end{tabular}
\end{minipage}
\hfill
% ==================== C|AB ====================
\begin{minipage}[t]{0.325\textwidth}
\vspace{0pt}
\centering
\begin{tabular}{@{}l l r@{}}
\multicolumn{3}{c}{\textbf{(c) bipartition $C|AB$}}\\
\toprule
$M$ & $P\subseteq M^c$ & $\det$\\
\midrule
$\{4,5\}$ & $\{1,2,3,6,7,8,9,10,11\}$ & $12$\\
$\{4,7\}$ & $\{1,2,3,5,6,8,9,11,12\}$ & $270$\\
$\{4,10\}$ & $\{1,2,3,5,6,7,8,9,11\}$ & $36$\\
$\{5,6\}$ & $\{1,2,3,4,7,8,9,10,11\}$ & $22$\\
$\{5,9\}$ & $\{1,2,3,4,6,7,8,10,11\}$ & $-39$\\
$\{6,7\}$ & $\{1,2,3,4,5,8,9,11,12\}$ & $-180$\\
$\{6,10\}$ & $\{1,2,3,4,5,7,8,9,11\}$ & $-66$\\
$\{7,9\}$ & $\{1,2,3,4,5,6,8,10,11\}$ & $90$\\
$\{9,10\}$ & $\{1,2,3,4,5,6,7,8,11\}$ & $72$\\
$\{1,7,14\}$ & $\{2,3,4,5,6,8,9,10,11\}$ & $-60$\\
$\{2,3,4\}$ & $\{1,5,6,7,8,9,10,11,12\}$ & $-228$\\
$\{2,3,5\}$ & $\{1,4,6,7,8,9,10,11,12\}$ & $76$\\
$\{2,3,6\}$ & $\{1,4,5,7,8,9,10,11,12\}$ & $180$\\
$\{2,3,7\}$ & $\{1,4,5,6,8,9,10,11,12\}$ & $60$\\
$\{6,8,13\}$ & $\{1,2,3,4,5,7,9,10,11\}$ & $110$\\
$\{6,11,12\}$ & $\{1,2,3,4,5,7,8,9,13\}$ & $84$\\
$\{7,11,12\}$ & $\{1,2,3,4,5,6,8,9,13\}$ & $252$\\
$\{9,11,12\}$ & $\{1,2,3,4,5,6,7,8,10\}$ & $-36$\\
$\{10,11,12\}$ & $\{1,2,3,4,5,6,7,8,9\}$ & $36$\\
$\{1,8,13,14\}$ & $\{2,3,4,5,6,7,9,10,11\}$ & $60$\\
$\{2,3,11,12\}$ & $\{1,4,5,6,7,8,9,10,13\}$ & $-224$\\
$\{1,2,3,10,14\}$ & $\{4,5,6,7,8,9,11,12,13\}$ & $576$\\
$\{1,4,6,9,14\}$ & $\{2,3,5,7,8,10,11,12,13\}$ & $1080$\\
$\{1,5,11,12,14\}$ & $\{2,3,4,6,7,8,9,10,13\}$ & $-60$\\
$\{2,3,8,9,13\}$ & $\{1,4,5,6,7,10,11,12,14\}$ & $-192$\\
$\{4,8,11,12,13\}$ & $\{1,2,3,5,6,7,9,10,14\}$ & $-60$\\
$\{5,7,8,10,13\}$ & $\{1,2,3,4,6,9,11,12,14\}$ & $120$\\
\bottomrule
\end{tabular}
\end{minipage}
\end{table*}

Direct integer minors show that the rank of every six vectors on each local party
is $3$: all $3\binom{14}{6}=9009$ cases pass.  Hence a locally
deficient $J$ in Eq.~\eqref{eq:criterion} has $|J|\leq 5$. Exhaustive
exact enumeration gives 228, 228, and 222 such subsets for $A|BC$,
$B|AC$, and $C|AB$, respectively (See Table~\ref{tab:app-deficient}). 
It suffices to consider only the inclusion-maximal subsets. Indeed, if
$J\subseteq M$, then $M^c\subseteq J^c$. Therefore,
$\operatorname{rank}
\{\ket{b_j}\otimes\ket{c_j}:j\in M^c\}=9$
implies that
$\operatorname{rank}
\{\ket{b_j}\otimes\ket{c_j}:j\in J^c\}=9$.
The cardinality distribution of the inclusion-maximal deficient subsets
is given in Table~\ref{tab:inclusion-maximal}. The bipartitions $A|BC$, $B|AC$,
and $C|AB$ have only $23$, $23$, and $27$ inclusion-maximal deficient
subsets, respectively. For each such subset $M$, the nine pair vectors
indexed by $M^c$, when arranged as the columns of an integer
$9\times9$ matrix, have a nonzero determinant. The complete determinant
certificates are listed in Table~\ref{tab:app-determinants}. It follows
from the inclusion-maximality argument above that, for every deficient
subset $J$ associated with each bipartition, the pair vectors indexed by $J^c$
have rank $9$.

Therefore, by Lemma~\ref{lemma:criterion},
$\{\ket{a_i}\otimes(\ket{b_i}\otimes\ket{c_i})\}_{i\in[14]}$,
$\{\ket{b_i}\otimes(\ket{a_i}\otimes\ket{c_i})\}_{i\in[14]}$, and
$\{\ket{c_i}\otimes(\ket{a_i}\otimes\ket{b_i})\}_{i\in[14]}$ are
bipartite UPBs in $\bbC^3\otimes\bbC^9$ across the bipartitions $A|BC$, $B|AC$,
and $C|AB$, respectively. Consequently,
$\{\ket{a_i}\otimes\ket{b_i}\otimes\ket{c_i}\}_{i\in[14]}$ is a GUPB.
See also End matter for the direct verification code in MATLAB.

% Thus for every cut  and every def
% Every determinant is a nonzero integer and therefore certifies full
% rank simultaneously over $\mathbb Q$, $\mathbb R$, and $\CC$.  For a
% nonmaximal deficient $J$, choose a maximal deficient $M\supseteq J$.
% The certified nine columns in $M^c$ are also present in $J^c$, so the
% second alternative in Eq.~\eqref{eq:rank-alternative} follows by
% monotonicity.  Hence the criterion fails for no subset on any cut.
% There is no product extension on $A|BC$, $B|AC$, or $C|AB$.
% Together with Eq.~\eqref{eq:orthogonality} and $14<27$, this proves
% that the states form a GUPB.

% The certificate was recomputed by three exact-arithmetic paths: the
% primary fraction-free integer verifier, an independent rational SymPy
% audit, and a table-only SymPy verifier.  All three return zero
% uncovered pairs, zero six-subset failures, and zero complement-rank
% violations.  The first checks all 678 deficient subsets, while the
% other two reconstruct the ranks independently; details and file
% hashes appear in Appendix~\hyperref[app:exact-audits]{E}.

% The thirteen-state impossibility theorem of
% Ref.~\cite{Demianowicz2026}, together with the lower bound of
% Ref.~\cite{Shi2023}, excludes every smaller three-qutrit candidate.
% The construction above therefore completes the proof of Theorem~1.

\textit{\textbf{GUPBs in arbitrary
tripartite local dimensions.}}\textbf{---}
The construction of the three-qutrit GUPB extends to arbitrary
tripartite local dimensions. Consider
$\bbC^{d_A}\otimes\bbC^{d_B}\otimes\bbC^{d_C}$, where
$3\leq d_A\leq d_B\leq d_C$. For $X=A,B,C$, define
$U_X=\operatorname{span}\{\ket{0},\ket{1},\ket{2}\}
\subset\bbC^{d_X}$, and let
$U=U_A\otimes U_B\otimes U_C$. Embed the three-qutrit GUPB
$\mathcal G_3=
\{\ket{a_i}\otimes\ket{b_i}\otimes\ket{c_i}\}_{i\in[14]}$
into $U$, and denote its image by $\mathcal G_3'$.

Let $\mathcal B_{\rm sh}$ consist of all computational-basis product
states $\ket{p,q,r}$ satisfying $0\leq p<d_A$, $0\leq q<d_B$, and
$0\leq r<d_C$, for which $\max\{p,q,r\}\geq3$. Define
\begin{equation}
\widetilde{\mathcal G}
=\mathcal G_3'\cup\mathcal B_{\rm sh},
\qquad
|\widetilde{\mathcal G}|=d_Ad_Bd_C-13.
\label{eq:dimension-extension}
\end{equation}
Indeed, $\mathcal B_{\rm sh}$ is an orthonormal basis of $U^\perp$ and
contains $d_Ad_Bd_C-27$ states. Hence,
$\widetilde{\mathcal G}^{\perp}$ is precisely the embedded copy of
$\mathcal G_3^\perp$. Since the embedding into $U$ is the tensor
product of three local embeddings, it preserves the product structure
across each of the bipartitions $A|BC$, $B|AC$, and $C|AB$. Therefore, if
$\widetilde{\mathcal G}^{\perp}$ contained a nonzero product vector
across any of these bipartitions, pulling it back to $(\bbC^3)^{\otimes3}$
would yield a nonzero product vector in $\mathcal G_3^\perp$ across
the same bipartition, contradicting the GUPB property of $\mathcal G_3$.
Thus, $\widetilde{\mathcal G}$ is a GUPB in
$\bbC^{d_A}\otimes\bbC^{d_B}\otimes\bbC^{d_C}$.

\textit{\textbf{Application 1: bound entangled states across every bipartition.}}\textbf{---}
Let $\mathcal G_3=\{\ket{\psi_i}\}_{i\in[14]}$ denote the normalized
version of the three-qutrit GUPB
$\{\ket{a_i}\otimes\ket{b_i}\otimes\ket{c_i}\}_{i\in[14]}$.
Since $\mathcal G_3$ is a GUPB, its orthogonal complement
$\mathcal G_3^\perp$ is a $13$-dimensional genuinely entangled
subspace; that is, every nonzero vector in $\mathcal G_3^\perp$ is
genuinely entangled. The normalized projector onto this subspace is
\begin{equation}
\rho=\frac{1}{13}\left(
I_{27}-\sum_{i=1}^{14}
\ket{\psi_i}\!\bra{\psi_i}
\right).
\label{eq:rho}
\end{equation}
Since the range of $\rho$ is $\mathcal G_3^\perp$, $\rho$ msut be genuinely entangled.

Moreover, every local vector in the construction has real coefficients
in the computational basis. Hence, each rank-one projector
$\ket{\psi_i}\!\bra{\psi_i}$ in Eq.~\eqref{eq:rho} is invariant under
partial transposition with respect to any subsystem. Consequently,
\begin{equation}
\rho^{T_A}=\rho^{T_B}=\rho^{T_C}=\rho.
\label{eq:ppt-invariant}
\end{equation}
Thus, $\rho$ is PPT and entangled across each of the bipartitions
$A|BC$, $B|AC$, and $C|AB$. Since PPT entangled states are
nondistillable, $\rho$ is bound entangled across every bipartition
\cite{Bennett1999,Horodecki1998}.

\textit{\textbf{Application 2: strong quantum nonlocality.}}\textbf{---}
A set of orthogonal product states is locally irreducible if no party
can eliminate one or more states by performing a nontrivial
orthogonality-preserving local measurement. Such a set is strongly
nonlocal if it remains locally irreducible across every bipartition,
thereby exhibiting strong quantum nonlocality without entanglement
\cite{Halder2019}.

Let
$\mathcal S=\{\ket{x_i}_A\otimes\ket{y_i}_B\}_{i\in[k]}
\subset\bbC^{d_A}\otimes\bbC^{d_B}$
be a set of orthogonal product states. Consider a local measurement
$\{M_m\}_m$ on subsystem $A$, with POVM elements
$E_m=M_m^\dagger M_m$ satisfying $\sum_m E_m=I_A$. The measurement is
orthogonality preserving if, for every outcome $m$, the corresponding
postmeasurement states remain mutually orthogonal. Equivalently, for
all $i\neq j$ and all $m$,
\begin{equation}
\bra{x_i}E_m\ket{x_j}_A\braket{y_i}{y_j}_B=0.
\label{eq:op-measurement}
\end{equation}
The measurement is trivial if every $E_m$ is proportional to $I_A$;
otherwise, it is nontrivial. The analogous definition applies to a
measurement on subsystem $B$.

We recall a useful criterion for ruling out nontrivial
orthogonality-preserving measurements on subsystem $A$
\cite{Li2023boundssmallestsets}. Define
\begin{equation}
\Gamma_A=
\{(i,j):i\neq j,\ 
\braket{x_i}{x_j}_A=0,\ 
\braket{y_i}{y_j}_B\neq0\}
\label{eq:gamma-A}
\end{equation}
and the associated operator space
\begin{equation}
\mathcal L_A=
\operatorname{span}
\{\ket{x_i}_A\!\bra{x_j}:(i,j)\in\Gamma_A\}.
\label{eq:operator-space-A}
\end{equation}
If $\dim\mathcal L_A=d_A^2-1$, then every
orthogonality-preserving local measurement on subsystem $A$ is
trivial. Indeed, Eq.~\eqref{eq:op-measurement} implies that every POVM
element $E_m$ on $A$ is Hilbert--Schmidt orthogonal to
$\mathcal L_A$. Moreover, every operator in $\mathcal L_A$ is
traceless. Therefore, the condition
$\dim\mathcal L_A=d_A^2-1$ implies that $\mathcal L_A$ is the full
traceless operator space on $\bbC^{d_A}$. Its Hilbert--Schmidt
orthogonal complement is $\operatorname{span}\{I_A\}$, and hence every
$E_m$ is proportional to $I_A$.

Next, we show that
$\mathcal G_3=
\{\ket{a_i}\otimes\ket{b_i}\otimes\ket{c_i}\}_{i\in[14]}$ is strongly nonlocal.
It is sufficient to prove that every orthogonality-preserving
measurement on each of the joint subsystems $AB$, $AC$, and $BC$ is
trivial \cite{Shi2022stronglynonlocal}. To treat the three cases uniformly, set
$\ket{u_i^{(A)}}=\ket{a_i}$,
$\ket{u_i^{(B)}}=\ket{b_i}$, and
$\ket{u_i^{(C)}}=\ket{c_i}$. For each $X\in\{A,B,C\}$, let $Y$ and $Z$
be the other two parties and define
$\ket{w_i^{(X)}}=
\ket{u_i^{(Y)}}\otimes\ket{u_i^{(Z)}}\in\bbC^9$.
Introduce the ordered-pair set
\begin{equation}
J_X=
\{(i,j):i\neq j,\ 
\braket{u_i^{(X)}}{u_j^{(X)}}\neq0,\ 
\braket{w_i^{(X)}}{w_j^{(X)}}=0\}.
\label{eq:JX}
\end{equation}
Using the integer vectors in Table~\ref{tab:construction}, an exact
calculation gives
\begin{equation}
\dim\operatorname{span}
\{\ket{w_i^{(X)}}\!\bra{w_j^{(X)}}:(i,j)\in J_X\}
=80=9^2-1
\label{eq:strong-nonlocal-rank}
\end{equation}
for every $X\in\{A,B,C\}$. 
For the bipartition $X|YZ$, Eq.~\eqref{eq:strong-nonlocal-rank} and
the operator-span criterion imply that every
orthogonality-preserving measurement on $YZ$ is trivial. Taking
$X=A,B,C$ rules out nontrivial measurements on $BC$, $AC$, and $AB$,
respectively. Thus the three-qutrit GUPB $\mathcal G_3$ exhibits
strong quantum nonlocality without entanglement. Note that strongly nonlocal UPBs that are not GUPBs also exist
\cite{Shi2022stronglynonlocal}.

\textit{\textbf{Conclusion and outlook.}}\textbf{---}
We have established the existence of GUPBs by constructing a
three-qutrit GUPB of cardinality fourteen. The three-qutrit system is
the smallest tripartite Hilbert space in which a GUPB can exist, so
this construction resolves a long-standing open question. Moreover,
fourteen is the minimum possible cardinality of a three-qutrit GUPB.
We have also extended the construction to arbitrary tripartite local
dimensions. As applications, the constructed GUPB gives rise to a
state that is PPT and bound entangled across every bipartition, and it
also exhibits strong quantum nonlocality without entanglement. It
would be interesting to construct minimum-cardinality GUPBs in other
multipartite Hilbert spaces.

\textit{\textbf{Acknowledgments.}}\textbf{---}
The construction of the three-qutrit GUPB was developed with the assistance of GPT-5.6.
The authors independently write the entire
manuscript, and they take full responsibility for the correctness of
the results and the content of the manuscript.

\textit{\textbf{Note added.}}\textbf{---}
While finalizing this manuscript, we became aware of independent
concurrent work constructing families of GUPBs from MDS codes for an
arbitrary number of parties $N\geq3$ \cite{Li2026MDS}. The smallest
explicit instance of that construction is a GUPB of cardinality
$34\,607$ in $(\bbC^{33})^{\otimes3}$, and its cardinality is not
claimed to be minimal. In contrast, the present work constructs a
 GUPB of cardinality 14 in $(\bbC^3)^{\otimes3}$, the smallest tripartite
Hilbert space admitting a GUPB, and proves that 14 is the minimum
possible cardinality in this system.

\begin{center}
 \textbf{End matter}
\end{center}
% \textit{\textbf{End matter.}}\textbf{---}
The following MATLAB script directly verifies the mutual orthogonality
of the fourteen product states and their unextendibility across the
three bipartitions $A|BC$, $B|AC$, and $C|AB$. It can be copied into
MATLAB and executed directly, without defining any functions or using
additional toolboxes.
\begin{lstlisting}[style=matlabcompact]
A = [
 0  1  0  0  1  1  1  1  0  1  1  1  1  0;
-1  0  2  1  2  0 -1  0  0  0 -1 -1  1  1;
 2  0 -1  0  1  0  1 -1  1 -1 -2  1  0  0];

B = [
 0  1  1  0  1  1  1  1  0  1  1  0  1  0;
 0 -1  0 -1  0 -1  0  0  1  1 -1  2  2  1;
 1 -2  0  2 -1  1  0 -1  0  0  1 -1  1  0];

C = [
 1  2  2  0  1  1  3  0  1 -1  0  0  0  1;
 0  1  1  1  0  3 -1  1  1  1  0  0  1  0;
 0 -1 -1  0  2  0  5  1  0 -1  1  1  1  0];

n = size(A,2);
tol = 1e-10;

G = (A'*A).*(B'*B).*(C'*C);
orthPass = all(triu(G,1) == 0,'all');
assert(orthPass,'Mutual orthogonality failed.');
fprintf('Mutual orthogonality: PASS\n');

BC = zeros(9,n);
AC = zeros(9,n);
AB = zeros(9,n);

for i = 1:n
    BC(:,i) = kron(B(:,i),C(:,i));
    AC(:,i) = kron(A(:,i),C(:,i));
    AB(:,i) = kron(A(:,i),B(:,i));
end

X = {A,B,C};
Y = {BC,AC,AB};
bipartitions = {'A|BC','B|AC','C|AB'};

for c = 1:3
    violation = false;

    for mask = 0:2^n-1
        J = logical(bitget(uint16(mask),1:n));

        if rank(X{c}(:,J),tol) < 3 && ...
           rank(Y{c}(:,~J),tol) < 9
            violation = true;
            break;
        end
    end

    assert(~violation,...
           ['Unextendibility failed on ',bipartitions{c}]);
    fprintf('%s unextendibility: PASS\n',bipartitions{c});
end

fprintf('FINAL RESULT: the states form a GUPB.\n');
\end{lstlisting}

\bibliography{reference}

@article{Bennett1999,
  author = {Bennett, Charles H. and DiVincenzo, David P. and Mor, Tal and Shor, Peter W. and Smolin, John A. and Terhal, Barbara M.},
  title = {Unextendible Product Bases and Bound Entanglement},
  journal = {Phys. Rev. Lett.},
  volume = {82},
  pages = {5385--5388},
  year = {1999},
  doi = {10.1103/PhysRevLett.82.5385}
}

@article{DiVincenzo2003,
  author = {DiVincenzo, David P. and Mor, Tal and Shor, Peter W. and Smolin, John A. and Terhal, Barbara M.},
  title = {Unextendible Product Bases, Uncompletable Product Bases and Bound Entanglement},
  journal = {Commun. Math. Phys.},
  volume = {238},
  pages = {379--410},
  year = {2003},
  doi = {10.1007/s00220-003-0877-6}
}

@article{Horodecki1998,
  author = {Horodecki, Micha{\l} and Horodecki, Pawe{\l} and Horodecki, Ryszard},
  title = {Mixed-State Entanglement and Distillation: Is There a ``Bound'' Entanglement in Nature?},
  journal = {Phys. Rev. Lett.},
  volume = {80},
  pages = {5239--5242},
  year = {1998},
  doi = {10.1103/PhysRevLett.80.5239}
}

@article{DemianowiczAugusiak2018,
  author = {Demianowicz, Maciej and Augusiak, Remigiusz},
  title = {From Unextendible Product Bases to Genuinely Entangled Subspaces},
  journal = {Phys. Rev. A},
  volume = {98},
  pages = {012313},
  year = {2018},
  doi = {10.1103/PhysRevA.98.012313}
}

@article{Shi2022,
  author = {Shi, Fei and Li, Mao-Sheng and Zhang, Xiande and Zhao, Qi},
  title = {Unextendible and Uncompletable Product Bases in Every Bipartition},
  journal = {New J. Phys.},
  volume = {24},
  pages = {113025},
  year = {2022},
  doi = {10.1088/1367-2630/ac9e14}
}

@article{Demianowicz2022,
  author = {Demianowicz, Maciej},
  title = {Negative Result about the Construction of Genuinely Entangled Subspaces from Unextendible Product Bases},
  journal = {Phys. Rev. A},
  volume = {106},
  pages = {012442},
  year = {2022},
  doi = {10.1103/PhysRevA.106.012442}
}

@article{Shi2023,
  author = {Shi, Fei and Bai, Ge and Zhang, Xiande and Zhao, Qi and Chiribella, Giulio},
  title = {Graph-Theoretic Characterization of Unextendible Product Bases},
  journal = {Phys. Rev. Research},
  volume = {5},
  pages = {033144},
  year = {2023},
  doi = {10.1103/PhysRevResearch.5.033144}
}

@article{Demianowicz2026,
  author = {Demianowicz, Maciej},
  title = {Progress in the Study of the (Non)Existence of Genuinely Unextendible Product Bases},
  journal = {Quantum Inf. Process.},
  volume = {25},
  pages = {67},
  year = {2026},
  doi = {10.1007/s11128-026-05072-w}
}

@article{Halder2019,
  title = {Strong Quantum Nonlocality without Entanglement},
  author = {Halder, Saronath and Banik, Manik and Agrawal, Sristy and Bandyopadhyay, Somshubhro},
  journal = {Phys. Rev. Lett.},
  volume = {122},
  issue = {4},
  pages = {040403},
  numpages = {7},
  year = {2019},
  month = {Feb},
  publisher = {American Physical Society},
  doi = {10.1103/PhysRevLett.122.040403},
  url = {https://link.aps.org/doi/10.1103/PhysRevLett.122.040403}
}

@article{Li2023boundssmallestsets,
  doi = {10.22331/q-2023-09-07-1101},
  url = {https://doi.org/10.22331/q-2023-09-07-1101},
  title = {Bounds on the smallest sets of quantum states with special quantum nonlocality},
  author = {Li, Mao-Sheng and Wang, Yan-Ling},
  journal = {{Quantum}},
  issn = {2521-327X},
  publisher = {{Verein zur F{\"{o}}rderung des Open Access Publizierens in den Quantenwissenschaften}},
  volume = {7},
  pages = {1101},
  month = sep,
  year = {2023}
}

@article{Shi2022stronglynonlocal,
  doi = {10.22331/q-2022-01-05-619},
  url = {https://doi.org/10.22331/q-2022-01-05-619},
  title = {Strongly nonlocal unextendible product bases do exist},
  author = {Shi, Fei and Li, Mao-Sheng and Hu, Mengyao and Chen, Lin and Yung, Man-Hong and Wang, Yan-Ling and Zhang, Xiande},
  journal = {{Quantum}},
  issn = {2521-327X},
  publisher = {{Verein zur F{\"{o}}rderung des Open Access Publizierens in den Quantenwissenschaften}},
  volume = {6},
  pages = {619},
  month = jan,
  year = {2022}
}

@article{Augusiak2011,
  title = {Bell Inequalities with No Quantum Violation and Unextendable Product Bases},
  author = {Augusiak, R. and Stasi\ifmmode \acute{n}\else \'{n}\fi{}ska, J. and Hadley, C. and Korbicz, J. K. and Lewenstein, M. and Ac\'{\i}n, A.},
  journal = {Phys. Rev. Lett.},
  volume = {107},
  issue = {7},
  pages = {070401},
  numpages = {5},
  year = {2011},
  month = {Aug},
  publisher = {American Physical Society},
  doi = {10.1103/PhysRevLett.107.070401},
  url = {https://link.aps.org/doi/10.1103/PhysRevLett.107.070401}
}

@article{Augusiak2012,
  title = {Tight Bell inequalities with no quantum violation from qubit unextendible product bases},
  author = {Augusiak, R. and Fritz, T. and Kotowski, Ma. and Kotowski, Mi. and Paw\l{}owski, M. and Lewenstein, M. and Ac\'{\i}n, A.},
  journal = {Phys. Rev. A},
  volume = {85},
  issue = {4},
  pages = {042113},
  numpages = {12},
  year = {2012},
  month = {Apr},
  publisher = {American Physical Society},
  doi = {10.1103/PhysRevA.85.042113},
  url = {https://link.aps.org/doi/10.1103/PhysRevA.85.042113}
}

@article{fritz2013local,
  title={Local orthogonality as a multipartite principle for quantum correlations},
  author={Fritz, Tobias and Sainz, Ana Bel{\'e}n and Augusiak, Remigiusz and Brask, J Bohr and Chaves, Rafael and Leverrier, Anthony and Ac{\'\i}n, Antonio},
  journal={Nat. Commun.},
  volume={4},
  number={1},
  pages={2263},
  year={2013},
  doi = {10.1038/ncomms3263},
  url = {https://doi.org/10.1038/ncomms3263},
  publisher={Nature Publishing Group UK London}
}

@article{Joh13,
	title={The Minimum Size of Qubit Unextendible Product Bases},
	author = {Johnston, Nathaniel},
	journal={Proceedings of the 8th Conference on the Theory of Quantum Computation, Communication and Cryptography (TQC 2013)},
	pages={93–105},
	volume={22},	
	Year={2013},
	organization={Schloss Dagstuhl–Leibniz-Zentrum fuer Informatik}
}

@article{Chen2013The,
  title={The minimum size of unextendible product bases in the bipartite case (and some multipartite cases)},
  author={Chen, Jianxin and Johnston, Nathaniel},
  journal={Commun. Math. Phys. },
  volume={333},
  number={1},
  pages={351--365},
  year={2015},
  url={https://doi.org/10.1007/s00220-014-2186-7},
  doi={10.1007/s00220-014-2186-7},
  publisher={Springer}
}

@article{AL01,
  title={Unextendible product bases},
  author={Alon, Noga and Lov{\'a}sz, L{\'a}szl{\'o}},
  journal={J. Combin. Theory, Ser. A},
  volume={95},
  number={1},
  pages={169--179},
  year={2001},
  doi = {https://doi.org/10.1006/jcta.2000.3122},
  url = {https://www.sciencedirect.com/science/article/pii/S0097316500931224},
  publisher={Citeseer}
}

@article{Fen06,
  title={Unextendible product bases and 1-factorization of complete graphs},
  author={Feng, Keqin},
  journal={Discrete Appl. Math.},
  volume={154},
  number={6},
  pages={942--949},
  year={2006},
  doi = {https://doi.org/10.1016/j.dam.2005.10.011},
url ={https://www.sciencedirect.com/science/article/pii/S0166218X05003513},
  publisher={Elsevier}
}

@article{Johnston2014The,
  title={The structure of qubit unextendible product bases},
  author={Johnston, Nathaniel},
  journal={J. Phys. A: Math. Theor.},
  volume={47},
  number={42},
  pages={424034},
  year={2014},
  doi = {10.1088/1751-8113/47/42/424034},
url = {https://doi.org/10.1088/1751-8113/47/42/424034},
  publisher={IOP Publishing}
}

@article{shi2020unextendible,
  title = {Unextendible product bases from tile structures and their local entanglement-assisted distinguishability},
  author = {Shi, Fei and Zhang, Xiande and Chen, Lin},
  journal = {Phys. Rev. A},
  volume = {101},
  issue = {6},
  pages = {062329},
  numpages = {9},
  year = {2020},
  month = {Jun},
  publisher = {American Physical Society},
  doi = {10.1103/PhysRevA.101.062329},
  url = {https://link.aps.org/doi/10.1103/PhysRevA.101.062329}
}

@article{Li2026MDS,
author= {Li, Mao-Sheng},
title= {Genuinely Unextendible Product Bases from Maximum Distance Separable Codes},
journal={arXiv:2608.09504},
year={2026},
url={https://doi.org/10.48550/arXiv.2608.09504},
}

\end{document}